\documentclass[aps,prl,amsmath,amssymb,reprint,superscriptaddress,]{revtex4-2}
\usepackage{charter,graphicx,verbatim,threeparttable,float,amssymb }
\usepackage[version=4]{mhchem}

\begin{document}

\title[CsV3Sb5-xSnx]{Fermi level tuning and double-dome superconductivity in the kagome metals CsV$_3$Sb$_{5-x}$Sn$_x$}

\author{Yuzki M. Oey}
\affiliation{Materials Department, Materials Research Laboratory, and California NanoSystems Institute\\
University of California Santa Barbara, California 93106 United States}
\email{yoey@ucsb.edu, stephendwilson@ucsb.edu}

\author{Brenden R. Ortiz}
\affiliation{Materials Department, Materials Research Laboratory, and California NanoSystems Institute\\
University of California Santa Barbara, California 93106 United States}

\author{Farnaz Kaboudvand}
\affiliation{Materials Department, Materials Research Laboratory, and California NanoSystems Institute\\
University of California Santa Barbara, California 93106 United States}

\author{Jonathan Frassineti}
\affiliation{Department of Physics, Brown University, Providence, RI 02912, U.S.A.}
\affiliation{Department of Physics and Astronomy ``A. Righi'', University of Bologna, I-40127 Bologna, Italy}
\author{Erick Garcia}
\affiliation{Department of Physics, Brown University, Providence, RI 02912, U.S.A.}
\author{Rong Cong}
\affiliation{Department of Physics, Brown University, Providence, RI 02912, U.S.A.}
\author{Samuele Sanna}
\affiliation{Department of Physics and Astronomy ``A. Righi'', University of Bologna, I-40127 Bologna, Italy}
\author{Vesna F. Mitrovi{\'c}}
\affiliation{Department of Physics, Brown University, Providence, RI 02912, U.S.A.}

\author{Ram Seshadri}
\affiliation{Materials Department, Materials Research Laboratory, and California NanoSystems Institute\\
University of California Santa Barbara, California 93106 United States}

\author{Stephen D. Wilson}
\affiliation{Materials Department, Materials Research Laboratory, and California NanoSystems Institute\\
University of California Santa Barbara, California 93106 United States}

\date{\today}

\begin{abstract}
The recently reported \textit{A}V$_3$Sb$_5$ (\textit{A} = K, Rb, Cs) family of kagome metals are candidates 
for unconventional superconductivity and chiral charge density wave (CDW) order; both potentially arise 
from nested saddle points in their band structures close to the Fermi energy. 
Here we use chemical substitution to introduce holes into CsV$_3$Sb$_{5}$ and unveil an unconventional
coupling of the CDW and superconducting states. Specifically, we generate a phase diagram for CsV$_3$Sb$_{5-x}$Sn$_{x}$ that illustrates 
the impact of hole-doping the system and lifting the nearest vHs toward and above $E_F$. Superconductivity exhibits a 
non-monotonic evolution with the introduction of holes, resulting in two ``domes'' peaked 
at 3.6\,K and 4.1\,K and the rapid suppression of three-dimensional CDW order. The evolution of CDW and superconducting 
order is compared with the evolution of the electronic band structure of CsV$_3$Sb$_{5-x}$Sn$_x$, where the complete suppression of superconductivity seemingly coincides with an electron-like band comprised of Sb $p_z$ orbitals pushed above E$_F$.
\end{abstract}

\maketitle

Kagome metals naturally support electronic structures 
that host Dirac points, flat bands, and saddle points, leading to electronic instabilities 
associated with divergences in the density of states at the Fermi level. A wide array of instabilities have been 
predicted, ranging from bond density wave order \cite{wang2013competing,isakov2006hard} to charge 
fractionalization \cite{obrien2010strongly, ruegg2011fractionally}, charge density waves 
(CDW) \cite{guo2009topological}, and superconductivity (SC) \cite{wang2013competing,ko2009doped}. As a result, the 
interplay between competing electronic instabilities can often be tuned \textit{via} small changes in the band filling. For example, band fillings near $5/4$ electrons per band \cite{wang2013competing,yu2012chiral,kiesel2013unconventional,barros2014exotic,feng2021chiral} can populate nested van Hove singularities (vHs) that drive CDW order, and, in some limits, unconventional superconductivity.

The recently discovered class of \textit{A}V$_3$Sb$_5$ (\textit{A}: K, Rb, Cs) kagome metals \cite{ortiz2019new} are 
potential realizations of this physical mechanism with each member exhibiting experimental signatures associated with CDW order \cite{jiang2021unconventional, zhao2021cascade, ortiz2020cs, ortiz2021superconductivity, liang2021three, chen2021roton} followed by the low temperature onset and coexistence of superconductivity \cite{ortiz2020cs, ortiz2021superconductivity, yin2021superconductivity}. Upon applying hydrostatic pressure, the CDW instability in \textit{A}V$_3$Sb$_5$ is coupled to superconductivity in a seemingly unconventional fashion \cite{du2021pressure, chen2021double}, and the impact of the pressure-modified band structure on the interplay between the two states remains an open area of study. In particular, understanding the relative roles of the vHs comprised of V $d$-orbital states near the $M$-points and the electron-like band comprised of Sb-states at the $\Gamma$-point within the CDW and SC mechanisms is essential to developing a microscopic picture of how the two transitions are coupled. 

Carrier doping is an appealing means of tuning these features relative to the Fermi level and probing the coupling of the CDW and SC states. Shifting the relative positions of the vHs and $\Gamma$ pocket relative to $E_F$ and probing the evolution and interplay of the CDW and SC phase transitions can provide insights into the origins of each state. A recent study on oxidized thin flakes of CsV$_3$Sb$_5$, for instance, shows that hole-doping on the Cs site can enhance $T_C$ while also suppressing CDW order \cite{song2021enhancement}. In addition, DFT calculations show that hole-doping drives the vHs in the opposite direction relative to E$_F$ than that expected \textit{via} external hydrostatic pressure \cite{labollita2021tuning}. Given that an unusual coupling between SC and CDW states was observed under variable pressure \cite{chen2021double}, a systematic study of hole-doping effects stands to provide an important experimental window into understanding this unconventional coupling.

Here the effect of hole-doping on the CDW and SC states in CsV$_3$Sb$_{5-x}$Sn$_x$ with 0 $\leq x\leq$ 1.5 is presented. Hole-doping is achieved \textit{via} substitution of Sn onto the Sb sites, and, because Sn and Sb are very similar in size, this drives negligible coincident steric effects in the band structure. As holes are introduced, the CDW state is rapidly suppressed, and three-dimensional CDW order vanishes near $x$ = 0.06. In parallel, SC is enhanced and reaches a maximum $T_C$=3.6\,K at $x$ = 0.03 within the CDW state before decreasing as the CDW is fully suppressed. Continued hole-doping beyond the suppression of CDW order results in a second maximum in $T_C$=4.1\,K at $x$ = 0.35 prior to bulk SC weakening and vanishing beyond $x$ = 0.5. Density functional theory (DFT) calculations and nuclear quadrupolar resonance measurements establish a strong preference for Sn to occupy Sb sites within the kagome plane, and DFT models predict that Sn substitution on this site lifts the Sb $p_z$ electron-like band at $\Gamma$ above E$_F$ coincident with the vanishing of bulk SC.  Furthermore, only small changes in the energies of the vHs are predicted for doping levels sufficient to suppress signatures of CDW order, suggesting the importance of Sb-states in the stabilization of both phases. 

\begin{figure}
\includegraphics[width=\linewidth]{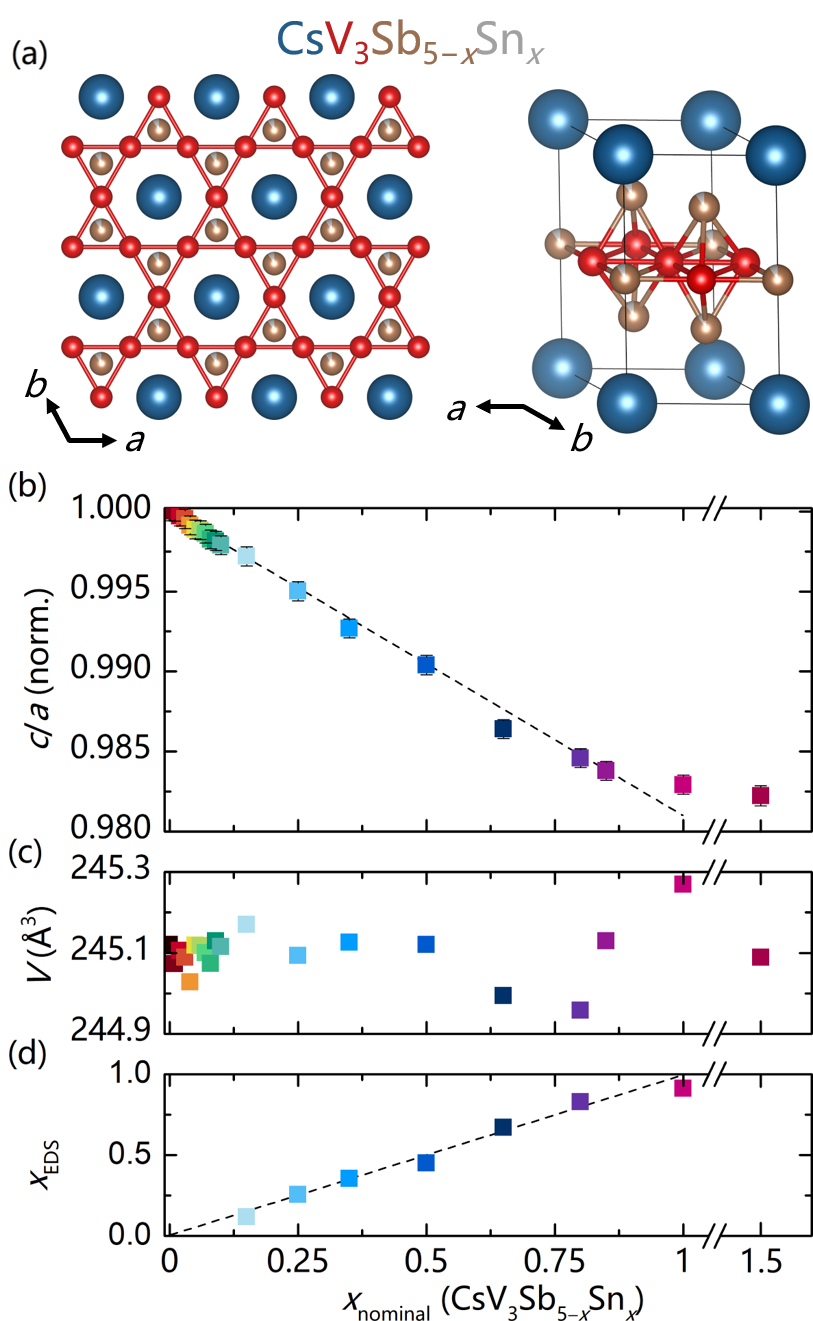}
\caption{(a) CsV$_3$Sb$_{5-x}$Sn$_x$ crystallizes in the parent CsV$_3$Sb$_5$ ($P$6/$mmm$) structure. Sn can substitute on either the Sb1 or Sb2 sublattices, and at low doping we establish a preference for the Sb site in the V kagome plane. (b-c) Sn integration into CsV$_3$Sb$_{5-x}$Sn$_x$ causes the c/a ratio to steadily decrease until the termination of solid solubility at $x$ = 1. (d) For samples with concentrations above EDS sensitivity threshold, nominal Sn tracks measured Sn. Error bars are shown unless they are smaller than associated point size. The different colors correspond to sample composition and are consistent throughout the figures.}
\label{fig:1structure}
\end{figure}

Powder samples of CsV$_3$Sb$_{5-x}$Sn$_x$ for 0 $\leq x \leq$ 1 in 6--7\,g batches were synthesized using a combination of ball milling and high-temperature sintering \cite{ESI}. Structural analysis was performed \textit{via} synchrotron powder x-ray data acquired at Argonne National Laboratory (APS, 11-BM) and on a Panalytical Empyrean laboratory x-ray powder diffractometer. A Hitachi TM4000Plus scanning electron microscope (SEM) was used to perform energy--dispersive x-ray spectroscopy (EDS). Magnetization data were collected in a Quantum Design Magnetic Property Measurement System (MPMS) and electrical resistivity data were collected in a Quantum Design Physical Property Measurement System (PPMS). DFT calculations were performed within the Vienna ab initio Simulation Package (VASP) \cite{kresse1996efficient,kresse1996efficiency}, and room temperature $^{121}$Sb zero-field NQR measurements were performed using a custom NMR spectrometer. Further details are available in the Supplemental Material accompanying this manuscript \cite{ESI}.

\begin{figure*}
\includegraphics[width=1\textwidth]{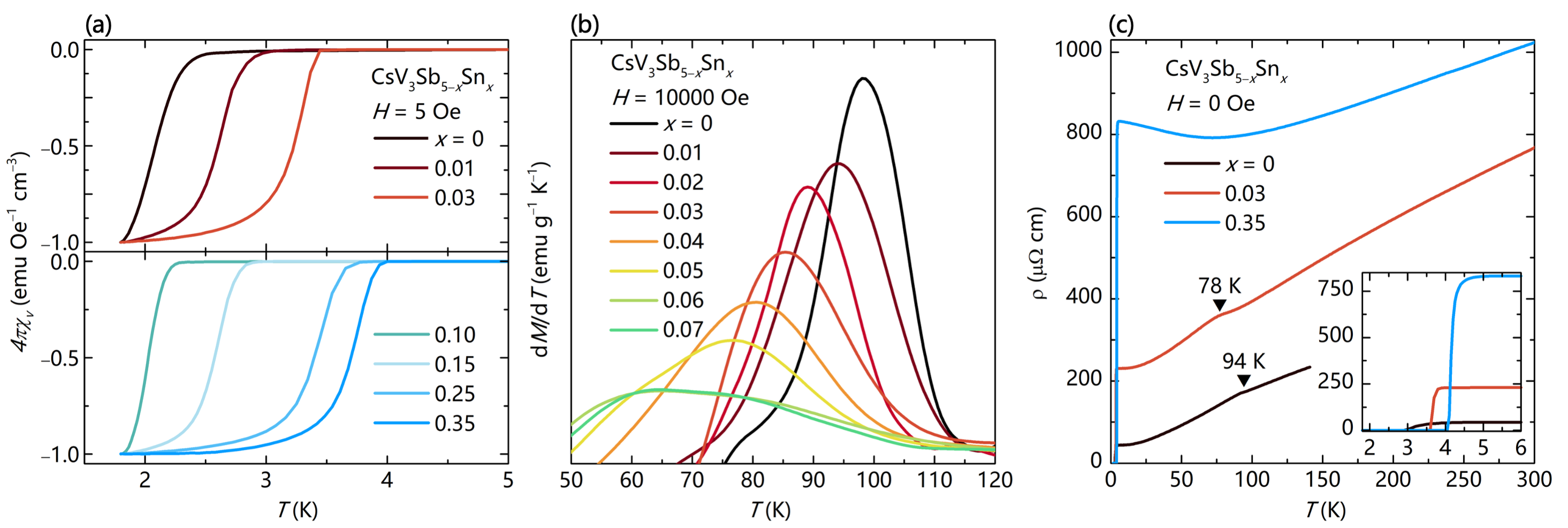}
\caption{(a) The superconducting $T_C$ measured under a field of $H$ = 5\,Oe shows a systematic shift to higher temperature in the compositions leading to the two $T_C$ maximums. The superconducting fraction is normalized to account for errors in mass and packing fraction so all data have a minimum of $-$1, the theoretical minimum. (b) d$M$/d$T$ for compositions $x \leq$ 0.06 show a decrease in CDW $T^*$, and this transition disappears for greater Sn compositions. (c) Resistivity data also show this enhancement in $T_C$, and the low-$T$, normal state resistivity for $x$ = 0.35 is about 4 times higher than $x$ = 0.03.}
\label{fig:2SCCDWdata}
\end{figure*}

\begin{figure}
\includegraphics[width=0.9\linewidth]{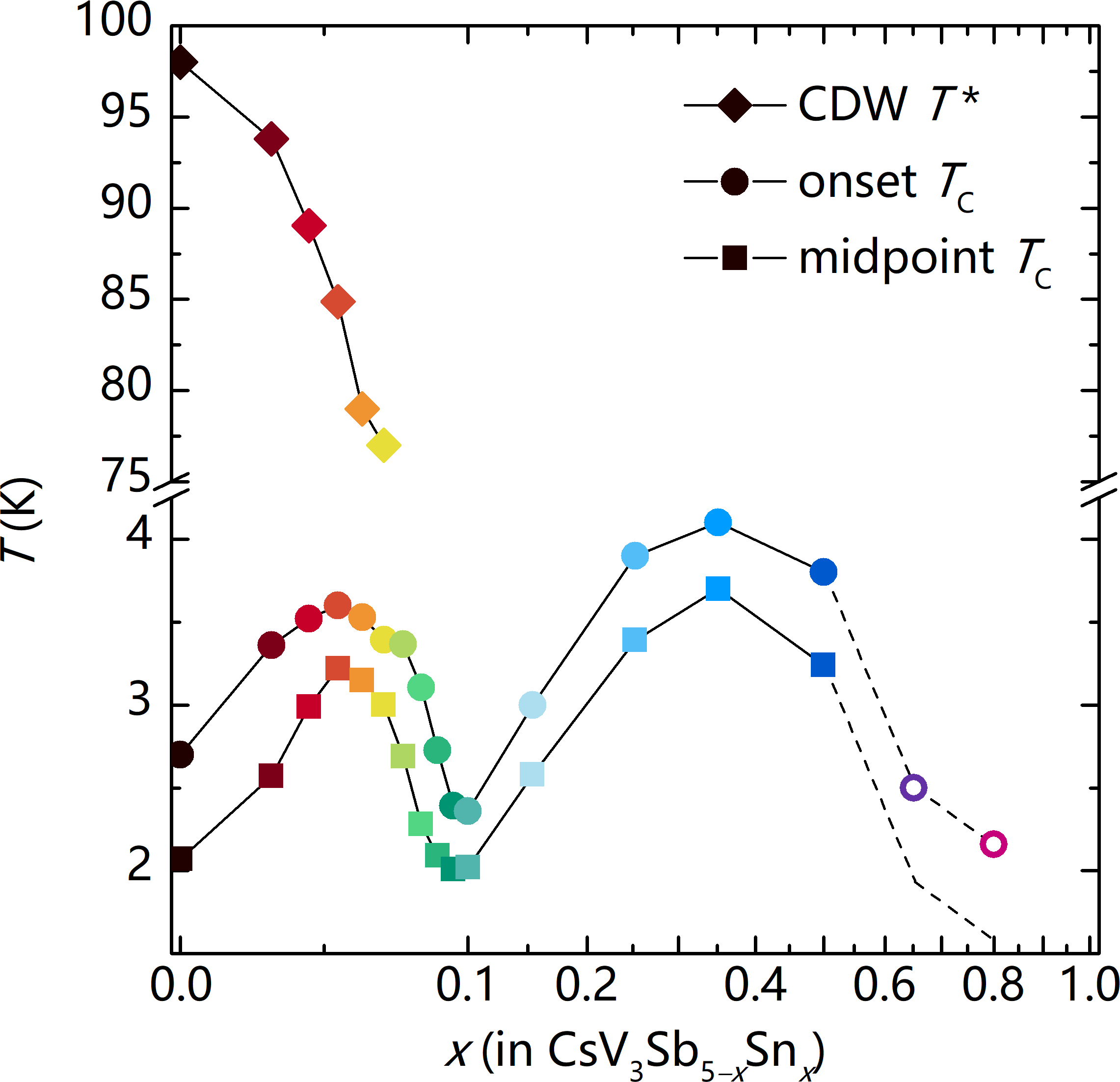}
\caption{Hole-doping phase diagram for CsV$_3$Sb$_{5-x}$Sn$_{x}$. The double-dome structure is clearly evident, as is the depression of the CDW ordering temperature. $T_C$ shows two maxima at $x$ = 0.03 and at $x$ = 0.35 and the CDW state disappears by $x$ = 0.06. We note that the first $T_C$ maximum occurs while the CDW is still present, and the CDW vanishes once $T_C$ is already declining again $x>$ 0.03. For $x \geq$ 0.65, the volume fraction of superconductivity decreases and the onset $T_C$s are represented with open circles.}
\label{fig:3phasediagram}
\end{figure}

At room temperature, CsV$_3$Sb$_{5-x}$Sn$_x$ with 0 $\leq x \leq$ 1 all adopt the same hexagonal $P6/mmm$ structure as CsV$_3$Sb$_5$ with an ideal kagome network of V atoms, shown in Fig.\,\ref{fig:1structure}(a).  For 0 $\leq x \leq$ 1, polycrystalline samples were found to be single phase, while a secondary phase appears for $x$ = 1.5, indicating the termination of the solid solution. Pawley fits of powder diffraction data were used to determine changes to the unit cell as a function of Sn content. Figures \ref{fig:1structure}(b-c) show the normalized $c/a$ ratio and the cell volume as a function of Sn content. With increasing Sn, $a$ increases and $c$ decreases in near perfect compensation, yielding a cell volume which is virtually independent of Sn content. Small fluctuations in volume of less than 0.16\% can be explained by uncertainties in chemical composition and in the structural refinement. 

The linear trend in $c/a$ is reminiscent of a Vegard's Law-type trend, suggesting that Sn is incorporated into the parent structure as a solid solution up to approximately $x$ = 1, at which point the $c/a$ ratio plateaus. Figure \ref{fig:1structure}(d) indicates that the measured Sn content (\textit{via} EDS) tracks the nominal Sn content. While Rietveld refinements of x-ray data confirm that powders are single phase, poor scattering contrast between Sb and Sn impedes the ability to refine whether Sn is uniformly distributed on both the Sb1 and Sb2 sites, or whether there is preferential occupation of a particular sublattice. DFT calculations suggest a preference $\approx$ 10\,meV/atom for Sn to substitute preferentially on the Sb1 (in the kagome plane) and nuclear quadropole resonance (NQR) measurements (discussed later) confirm that Sn atoms preferentially occupy this site \cite{ESI}.

Undoped CsV$_3$Sb$_5$ single crystals have a superconducting $T_C$=2.5\,K \cite{ortiz2020cs,song2021enhancement} and a charge density wave transition temperature $T^*$=94\,K \cite{zhao2021cascade,liang2021three} that drives a structural distortion into a $2\times 2\times \times 4$ enlarged unit cell; one containing layers of kagome nets distorted into both star-of-David and tri-hexagonal structures \cite{ortiz2021fermi,zhao2021cascade, hu2022coexistence}. Polycrystalline samples show identical transition temperatures, although both transitions are broader compared to single crystals \cite{ortiz2020cs}. This is largely due to powders being very sensitive to exact synthesis conditions and strain effects \cite{ESI}. Pure CsV$_3$Sb$_5$ powder synthesized here has an onset $T_C$ of 2.70\,K and a midpoint $T_C$ of 2.07\,K. As Sn is added within the matrix, Figure \ref{fig:2SCCDWdata}(a) shows the evolution of superconductivity, focusing on two different doping regimes ($x$ = 0 to 0.03 and $x$ = 0.1 to 0.35). All superconducting samples show $4 \pi \chi_v \approx -1$ in magnetization. Fractions slightly exceeding unity are attributed to errors in the packing density, and data are normalized to $-1$ for ease of comparison. The effect of Sn on $T_C$ becomes immediately apparent: as the Sn content is increased to $x$ = 0.03, $T_C$ increases to a local maxima of 3.6\,K. $T_C$ then decreases with continued hole-doping, followed by a second superconducting dome appearing shortly thereafter with a maximum $T_C$ = 4.1\,K at x=0.35. Superconductivity assumes only a partial volume fraction for $x >$ 0.5, and the SC state completely disappears by $x$ = 0.85. An exhaustive set of susceptibility measurements is provided in the supplemental information \cite{ESI}.

\begin{figure*}
\includegraphics[width=1\textwidth]{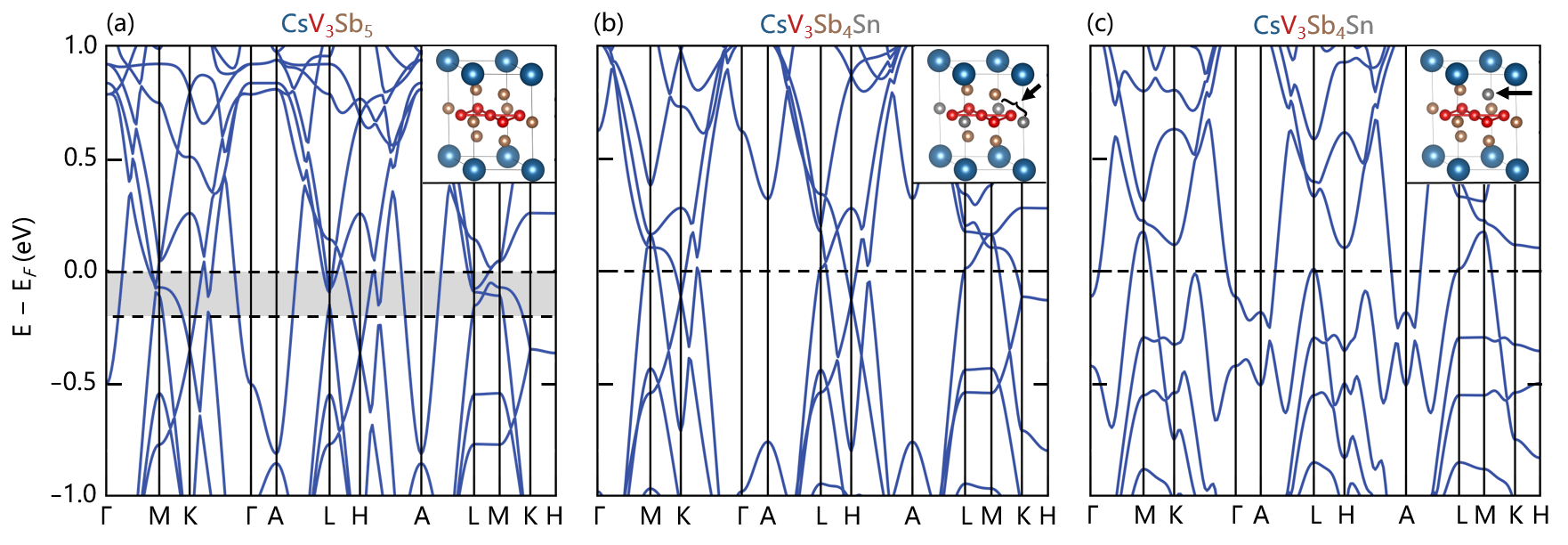}
\caption{(a) DFT calculations for CsV$_3$Sb$_5$ highlighting the allowable range of Fermi levels under the rigid band approximation for substitution of one Sn atom per formula unit. (b) Calculation of the band structure of CsV$_3$Sb$_4$Sn where one Sn has been substituted within the kagome plane. Here the majority of the electronic structure is preserved with the exception of the $\Gamma$ pocket, where the occupied Sb band is replaced with an unoccupied Sn band above $E_F$. (c) Calculation of the band structure of CsV$_3$Sb$_4$Sn where one Sn has been substituted at a Sb site outside of the kagome plane (OKP). A strong reconstruction of many bands can be observed, in particular near K, H, and K-$\Gamma$. }
\label{fig:4dft}
\end{figure*}

The evolution of CDW order upon hole-doping can be monitored \textit{via} via the inflection point in magnetization that appears at the CDW onset. Using the peak in d$M$/d$T$ as a metric for the onset of three-dimensional CDW order, Fig.\,\ref{fig:2SCCDWdata}(b) shows that light hole substitution causes the CDW transition to rapidly broaden and shift to lower temperatures. Undoped, the peak in d$M$/d$T$ is 98.01\,K and this peak decreases to less than 80\,K at x=0.04. For $x >$ 0.05, signatures of the CDW transition in magnetization cannot be resolved, although it is possible that a highly broadened CDW anomaly smoothly continues its rapid suppression with higher Sn doping. Despite the ambiguity in the precise position of CDW phase boundary near $x$=0.05, the superconducting and CDW states clearly coexist at the first maximum of the first superconducting dome ($x$ = 0.03). This is distinct from the second maximum in $T_C$ at higher Sn content ($x$ = 0.35), which occurs far beyond the apparent suppression of CDW order. Future work studying single crystals will be required to fully map if the suppression of the CDW state is a first or second order phase boundary and to fully explore whether CDW order is becoming short-ranged at larger Sn-doping levels.

Figure \ref{fig:2SCCDWdata}(c) shows electrical resistivity measurements performed on samples at both peak $T_C$ values ($x=$0.03 and $x=$0.35). Zero-resistivity conditions agree well with the $T_C$ obtained \textit{via} magnetization measurements, and the CDW onset temperature is further marked for the sample with $x$ = 0.03. The resistance immediately prior to the onset of SC increases quite dramatically upon Sn substitution, from 50\,$\mu\Omega$-cm ($x$ = 0), to 250\,$\mu\Omega$-cm ($x$ = 0.03, first $T_C$ peak), to 800\,$\mu\Omega$-cm ($x$ = 0.35, second $T_C$ peak), and the temperature dependence of the resistivity, particularly near the CDW temperature, changes dramatically with Sn doping.  As a summary, Figure \ref{fig:3phasediagram} plots a phase diagram showing the effect of Sn-substitution on both SC and CDW orders. A two-dome structure is immediately evident in $T_C$, as is the relatively rapid suppression of CDW order with the introduction of Sn.

 The CsV$_3$Sb$_5$ lattice supports nearly 1/5 of the total Sb being replaced by Sn. To give a rough sense of the range over which $E_F$ can be tuned, Figure \ref{fig:4dft}(a) shows the electronic structure of pure CsV$_3$Sb$_5$ along with the range of Fermi levels achievable with a loss of 1 electron per unit cell (gray shaded region). In a rigid band shift approximation, substitution of Sn should allow $E_F$ to be tuned across multiple vHs and Dirac points near the $M$- and $K$-points, respectively. However, reconstruction of the bands associated with the Sb orbitals and orbital selective doping effects are expected, and Figure \ref{fig:4dft}(c) shows one hypothetical structure with Sn substituted on the Sb2 sublattice (out of the kagome plane). Here, in addition to the expected shift of the Fermi level (note the position of the $M$-point vHs crossing now $\approx$ 0.2\,eV \textit{above} $E_F$), a significant reconstruction is also resolved around the K, L, and H points. Notably, the electron-like band at $\Gamma$ is preserved in this $x = 1$ structure. Alternatively, if Sn substitution instead occupies the Sb1 sublattice within the kagome plane, the overall electronic structure relative to CsV$_3$Sb$_5$ is largely preserved and mimics a rigid band shift model, with the important exception of the Sb-derived $\Gamma$-pocket, which vanishes and is replaced with an electron-like Sn-band far above $E_F$. Colorized orbital contributions to the band structure structures shown in Fig. 4 are presented in the supplemental information \cite{ESI}.
 
To clarify the impact of Sn-doping on the band structure, NQR measurements were performed and establish that Sn preferentially occupies the Sb1 sites in the kagome plane \cite{ESI}.  This is consistent with DFT calculations that show a small energy preference for Sn occupying the Sb1 site.  With this doping-site established, supercell calculations at fractional in-plane Sb substitution were performed and show that, as Sn content is increased, the electron-like band at $\Gamma$ lifts above $E_F$ near $x$ = 0.5 in tandem with the disappearance of bulk superconductivity \cite{ESI}. This suggests that superconductivity in CsV$_3$Sb$_5$ in-part relies on the Sb $p_z$-derived orbitals, and aligns with results of earlier pressure studies  \cite{tsirlin2021anisotropic}. 

At smaller Sn-doping levels, the observation of an intermediate peak in $T_C$ demonstrates a complex interplay between the Sb-states at $\Gamma$ and the V $d$-states driving the CDW order. Supercell calculations show a rather mild shift in the $M$-point vHs from roughly $-$0.072\,eV below $E_F$ for $x$=0 to $-$0.047\,eV for $x$=0.33 and $-$0.042\,eV for $x$=0.5, prior to shifting 0.083\,eV above $E_F$ in $x$=1. The small shift in the expected energies of the vHs toward $E_F$ under mild $x$=0.05 substitution fails to explain the rapid suppression of the CDW state at this doping, suggesting a complicated balance of energy scales underpinning the CDW state.  This is consistent with the rapid suppression of the CDW state under hydrostatic pressure \cite{du2021pressure, chen2021double}; however the two SC domes apparent under hole-doping seem distinct from those realized under pressure.  Superconductivity in crystals nominally doped between the SC domes retains a sharp transition \cite{ESI}, unlike the weak SC observed in pressure studies.  Further work exploring the distinctions between pressure (pushing vHs away from $E_F$) and hole-doping (pulling vHs toward $E_F$) is required to fully explore the parallels between the hole-doping and pressure-driven phase diagrams.

Hole-doping realized \textit{via} chemical substitution of Sn into CsV$_3$Sb$_5$ results in a complex electronic phase diagram featuring two SC domes -- one that coexists with CDW order and a second that appears following the suppression of the CDW state. Unconventional superconductivity has been predicted for fillings slightly away from the vHs \cite{kiesel2013unconventional}. The enhancement of $T_C$ prior to the suppression of CDW order and the subsequent second peak in $T_C$ following the complete suppression of the CDW state motivate deeper theoretical studies into how the Sb $p_z$ states intertwine with both CDW order SC in these materials. Our results demonstrate that small changes in the electronic structure achieved through carrier doping can have dramatic impacts on SC and CDW orders in CsV$_3$Sb$_5$ and provide a elegant chemical means to tune the competition between these states in the new \textit{A}V$_3$Sb$_5$ class of kagome superconductors. 

\section{acknowledgments}
This work was supported by the National Science Foundation (NSF) through Enabling Quantum Leap: Convergent
Accelerated Discovery Foundries for Quantum Materials Science, Engineering and Information (Q-AMASE-i): Quantum Foundry at UC Santa Barbara (DMR-1906325). The research reported here made use of shared facilities of the NSF Materials Research Science and Engineering Center at UC Santa Barbara DMR-1720256, a member of the Materials Research Facilities Network (www.mrfn.org). Use of the Advanced Photon Source at Argonne National Laboratory was supported by the U.S. Department of Energy, Office of Science, Office of Basic Energy Sciences, under Contract No. DE-AC02-06CH11357. YMO is supported by the National Science Foundation Graduate Research Fellowship Program under Grant No. DGE-1650114. BRO is supported by the California NanoSystems Institute through the Elings Fellowship program. FK acknowledges the Roy T. Eddleman Center for Quantum Innovation (ECQI) for their support. Work at Brown was supported in part by the the National Science Foundation grant No. DMR-1905532 and funds from Brown and University of Bologna.

\section*{References}
\bibliography{135frustratedmagnets}

\end{document}